\documentclass[twocolumn,showpacs,aps,prl,floats,superscriptaddress]{revtex4}

\usepackage{graphicx}
\usepackage{dcolumn}
\usepackage{amsmath}
\usepackage{epsfig}
\usepackage{verbatim}
\usepackage{color}
\usepackage{comment}
\input babarsym

\providecommand{\tbline}{\noalign{\vskip 0.04truecm\hrule\vskip0.04truecm}}
\def\leptontag{{\tt Lepton}\xspace}
\def\kaonitag{{\tt Kaon\,I}\xspace}
\def\kaoniitag{{\tt Kaon\,II}\xspace}
\def\kpitag{{\tt Kaon-Pion}\xspace}
\def\piontag{{\tt Pion}\xspace}
\def\othertag{{\tt Other}\xspace}

\def\nbb{\ensuremath{(383 \pm 4) \times 10^{\,\!6}\; \FourS \to \BB}\xspace}

\def\effectiveeta{0.504 \pm 0.033}

\def\fitr{0.233}\def\statr{0.010}\def\systr{0.005}
\def\measurerperp{\fitr \pm \statr\, \stat \pm \systr\, \syst}

\def\fitstwob{0.714}\def\statstwob{0.032}\def\syststwob{0.018}
\def\fitlambda{0.952}\def\statlambda{0.022}\def\systlambda{0.017}

\def\measures2twob{\ensuremath{\fitstwob \pm \statstwob \pm \syststwob \,\syst}\xspace}
\def\fitstwobstat{\ensuremath{\fitstwob \pm \statstwob}\xspace}

\def\fitlambdastat{\ensuremath{\fitlambda \pm \statlambda}\xspace}

\def\Imlambda{\ensuremath{\mathop{\cal I\mkern -2.0mu\mit m}\lambda}}
\def\abslambda{\ensuremath{|\lambda|}\xspace}
\def\Btag{\ensuremath{B_\mathrm{tag}}\xspace}
\def\Brec{\ensuremath{B_\mathrm{rec}}\xspace}
\def\BCP{\ensuremath{B_{CP}}\xspace}
\def\Bflav{\ensuremath{B_\mathrm{flav}}\xspace}
\def\deltaE{\ensuremath{\Delta E}\xspace}
\def\Gammad{\ensuremath{\Gamma_d}\xspace}
\def\deltaGammad{\ensuremath{\Delta \Gammad}\xspace}
\def\JpsiKsCh{\ensuremath{\jpsi\KS\,(\pipi)}\xspace}
\def\JpsiKszz{\ensuremath{\jpsi\KS\,(\ppz)}\xspace}
\def\Abar    {\kern 0.20em\overline{\kern -0.20em A}{}\xspace}
\def\Ab      {\ensuremath{\Abar}\xspace}

\def\etal{\textit{et al.}}
\newcommand{\BABARPubYear}     {07}
\newcommand{\BABARPubNumber}   {18}
\newcommand{\SLACPubNumber} {12391}

\begin{document}

\begin{flushleft}
\babar\texttt{\large -PUB-\BABARPubYear/\BABARPubNumber}\\
\texttt{\large SLAC-PUB-\SLACPubNumber} \\
\end{flushleft}

\title{\boldmath
Improved Measurement of \CP Violation in Neutral \B Decays to $\ccbar s$
}

%
\author{B.~Aubert}
\author{M.~Bona}
\author{D.~Boutigny}
\author{Y.~Karyotakis}
\author{J.~P.~Lees}
\author{V.~Poireau}
\author{X.~Prudent}
\author{V.~Tisserand}
\author{A.~Zghiche}
\affiliation{Laboratoire de Physique des Particules, IN2P3/CNRS et Universit\'e de Savoie, F-74941 Annecy-Le-Vieux, France }
\author{J.~Garra~Tico}
\author{E.~Grauges}
\affiliation{Universitat de Barcelona, Facultat de Fisica, Departament ECM, E-08028 Barcelona, Spain }
% ...
\collaboration{The \babar\ Collaboration}
\noaffiliation

\date{October 25, 2007}

\begin{abstract}
We present updated measurements of time-dependent $CP$ asymmetries in
fully-reconstructed neutral $B$ decays to several $CP$ eigenstates
containing a charmonium meson. The measurements use a data sample of
$(383 \pm 4) \times 10^{\,\!6}\; \Upsilon(4S) \to B\kern 0.18em\overline{\kern -0.18em B}{}$
decays collected with the \babar\ detector at the PEP-II $B$ factory.
We determine 
$\sin\!2\beta =  0.714 \pm 0.032 \,(\mathrm{stat}) \pm 0.018 \,(\mathrm{syst})$ and
$|\lambda| =  0.952 \pm 0.022 \,(\mathrm{stat}) \pm 0.017 \,(\mathrm{syst})$.
\end{abstract}


\maketitle

The standard model (SM) of electroweak interactions describes
\CP violation as a consequence of an
irreducible phase in the three-family Cabibbo-Kobayashi-Maskawa (CKM)
quark-mixing matrix~\cite{ref:CKM}.
In the CKM framework, neutral \B decays to \CP eigenstates containing a
charmonium and a $K^{(*)0}$ meson through tree-diagram dominated processes
provide a direct measurement of
\stwob~\cite{BCP}, where the angle $\beta$ is defined in terms of the CKM matrix
elements $V_{ij}$ as $\arg [-(V_{cd}^{}V_{cb}^*) / (V_{td}^{}V_{tb}^*)]$.

We report updated measurements, based on a sample of \nbb decays,
of \stwob and of the parameter \abslambda. Here
$\lambda=(q/p)(\Ab/A)$~\cite{ref:lambda}, 
$q$ and $p$ are complex constants that relate the \B-meson flavor
eigenstates to the mass eigenstates, and $\Ab/A$ is the ratio of
amplitudes of the decay of a \Bzb or \Bz to the final state under study.
We reconstruct \Bz decays to the final states
$\jpsi\KS$,
$\jpsi\KL$, $\psitwos\KS$, $\chicone\KS$, $\etac\KS$, and
$\jpsi\Kstarz$~\cite{ref:chargeconj}.
Since our previously published result~\cite{ref:babar2004}, we have added
$157\times10^6\;\BB$ decays and applied improved event reconstruction algorithms
to the entire dataset. We have also developed a new $\etac\KS$ event
selection based on the Dalitz plot structure of the $\etac\to\KS\Kp\pim$
decay, and have performed a more detailed study of the \CP
properties of the background events, which results in reduced systematic errors.
We now include the $\jpsi\KL$ and $\jpsi\Kstarz$ modes in
the sample to measure \abslambda, and we report individual measurements of
\stwob and \abslambda for each of the \CP decay modes used in the analysis.
Finally, we present separate results for the $\jpsi\KS$($\pipi+\ppz$)
\cite{ref:grossman}, and $\jpsi\Kz$($\KS+\KL$) modes.

We identify (tag) the initial flavor of the reconstructed \B
candidate, \Brec, using information from the other \B meson, \Btag, in the
event.
The decay rate $g_+$ $(g_-)$ for a neutral \B meson decaying to a \CP
eigenstate accompanied by a \Bz (\Bzb) tag can be expressed as
\begin{eqnarray}
&&g_\pm(\deltat) = \frac{e^{{- \left| \deltat \right|}/\tau_{\Bz} }}{4\tau_{\Bz} }
\Bigg\{ (1\mp\Delta\mistag) \Bigg.  \pm  (1-2\mistag)
\times  \nonumber \\
&&\Big[{\frac{2\Imlambda}{1+\abslambda^2}}\sin(\deltamd\deltat) -
 \Bigg. {\frac{1 - \abslambda^2 } {1 + \abslambda^2} }
       \cos(\deltamd\deltat)  \Big] \Bigg\}\:\:\:
\label{eq:timedist}
\end{eqnarray}
where $\deltat \equiv t_\mathrm{rec} - t_\mathrm{tag}$ is the difference
between the proper decay times of the reconstructed and tag \B mesons,
$\tau_{\Bz}$ is the neutral \B lifetime and \deltamd is the mass difference
of the \B meson mass eigenstates
determined from $\Bz-\Bzb$ oscillations~\cite{ref:pdg2006}. We assume that
the corresponding decay-width difference \deltaGammad is zero.
The average mistag probability \mistag describes the effect of incorrect
tags, and $\Delta\mistag$ is the difference between the mistag probabilities
for \Bz and \Bzb.
The sine term in Eq.~\ref{eq:timedist} results from the interference
between direct decay and decay after $\Bz-\Bzb$ oscillation. A
non-zero cosine term arises from the interference between decay amplitudes
with different weak and strong phases (direct \CP violation) or from \CP
violation in $\Bz-\Bzb$ mixing.
In the SM, \CP violation in mixing and
direct \CP violation in 
$b \to \ccbar s$ decays are both
negligible~\cite{ref:lambda}. Under these assumptions, $\lambda=\eta_f
e^{-2i\beta}$, where $\eta_f=\pm 1$ is the \CP eigenvalue of the final state
$f$. Thus, the time-dependent \CP-violating asymmetry is
\begin{eqnarray}
A_{\CP}(\deltat) &\equiv& \frac{g_+(\deltat) - g_-(\deltat)}{g_+(\deltat) +
g_-(\deltat)} \\ \nonumber
&=& -(1-2\mistag)\eta_f \stwob \sin{ (\deltamd \, \deltat )}.
\label{eq:asymmetry}
\end{eqnarray}

The \babar\ detector is described in detail elsewhere \cite{ref:babar}.
We select a sample of neutral \B mesons (\BCP) decaying to the $\eta_f=-1$
final states $\jpsi\KS$, $\psitwos\KS$, $\chicone\KS$ and 
$\etac \KS$, and to the $\eta_f=+1$ final state $\jpsi\KL$.
We reconstruct $\KS\to\pipi$, except in $\jpsi\KS$, where we also include
$\KS\to\piz\piz$.
The charmonium mesons are reconstructed in the
decays $\jpsi\to\epem,\,\mumu$; $\psitwos\to\epem,\,\mumu,\,\jpsi\pipi$;
$\chicone\to\jpsi\gamma$ and $\etac\to\KS\Kp\pim$.
We also reconstruct the $\jpsi\Kstarz (\Kstarz \to \KS\piz)$ 
final state, which can be \CP-even or \CP-odd due to the presence of even
($L$=0, 2) and odd ($L$=1) orbital angular momentum contributions.
Ignoring the angular information in $\jpsi\Kstarz$ results in a dilution
of the measured \CP asymmetry by a factor $\vert 1-2R_{\perp} \vert$, where
$R_{\perp}$ is the fraction of the $L$=1 contribution.
In Ref.~\cite{ref:rperp} we have measured $R_{\perp} = \measurerperp$,
which gives an effective $\eta_f = \effectiveeta$ for $f=\jpsi\Kstarz$,
after acceptance corrections.

In addition to the \CP modes described above, we use a sample of 
\Bz mesons (\Bflav) decaying to the flavor eigenstates
$D^{(*)-}h^+$ $(h^+=\pip,\rho^+,a_1^+)$ and
$\jpsi\Kstarz\;(\Kstarz\to\Kp\pim)$ to 
calibrate the flavor-tagging performance and \deltat resolution.
We also perform studies to measure apparent \CP violation
arising from \CP-conserving processes
using a control sample of \Bp mesons decaying to the final states
$\jpsi K^{(*)+}$, $\psitwos \Kp$, $\chicone \Kp$, and $\etac \Kp$.
The event selection and candidate reconstruction remain unchanged from those
described in Refs.~\cite{ref:babar2004,ref:bigprd,ref:etacks}, with the
exception of modes containing \etac mesons. In Ref.~\cite{ref:babar2004} we
reconstructed the $\Bz \to \etac \KS$ and $\Bpm \to \etac \Kpm$ modes using
the $\etac \to \KS \Kp \pim$ decay, with the requirement
$2.91 < m_{\KS \Kp \pim} < 3.05 \gevcc$.
We now exploit the fact that the \etac decays predominantly through a
$K\pi$ resonance at around $1430\mevcc$ and a $\KS K$ resonance close to
threshold, and require that either $m_{\KS\pim}$ or $m_{\Kp\pim}$ be in
the mass-range $[1.26,1.63]\gevcc$, or that $m_{\Kp\KS} \in [1.0,1.4] \gevcc$.

We calculate the time interval \deltat between the two \B decays from the
measured separation \deltaz between the decay vertices of \Brec and \Btag
along the collision ($z$) axis \cite{ref:bigprd}. 
The $z$ position of the \Brec vertex is determined from the charged
daughter tracks. The \Btag decay vertex is determined by fitting tracks not
belonging to the \Brec candidate to a common vertex,
while employing constraints from the beamspot location and the
\Brec momentum~\cite{ref:bigprd}. 
Events are accepted if the calculated $\deltat$ uncertainty is less than
$2.5\ps$ and $|\deltat|$ is less than $20\ps$. The fraction of all
events satisfying these requirements is $95\,\%$.

The algorithm used to determine the flavor of the \Btag at its decay to be
either \Bz or \Bzb is described in detail in Ref.~\cite{ref:babar2004}.
In brief, we define six mutually exclusive tagging
categories in order of decreasing tag purity: 
\leptontag, \kaonitag, \kaoniitag, \kpitag, \piontag and \othertag.
The figure-of-merit for tagging is the effective tagging efficiency
$Q\equiv \sum_i {\eps_i (1-2\mistag_i)^2}$, where $\eps_i$ is the
tagging efficiency of tagging category $i$.
We measure $Q = (30.5 \pm 0.3)\,\%$, consistent with the results
in Ref.~\cite{ref:babar2004}.

\begin{figure}[!htb]
\begin{center}
\includegraphics[bb=0 0 567 520,width=8.5cm]{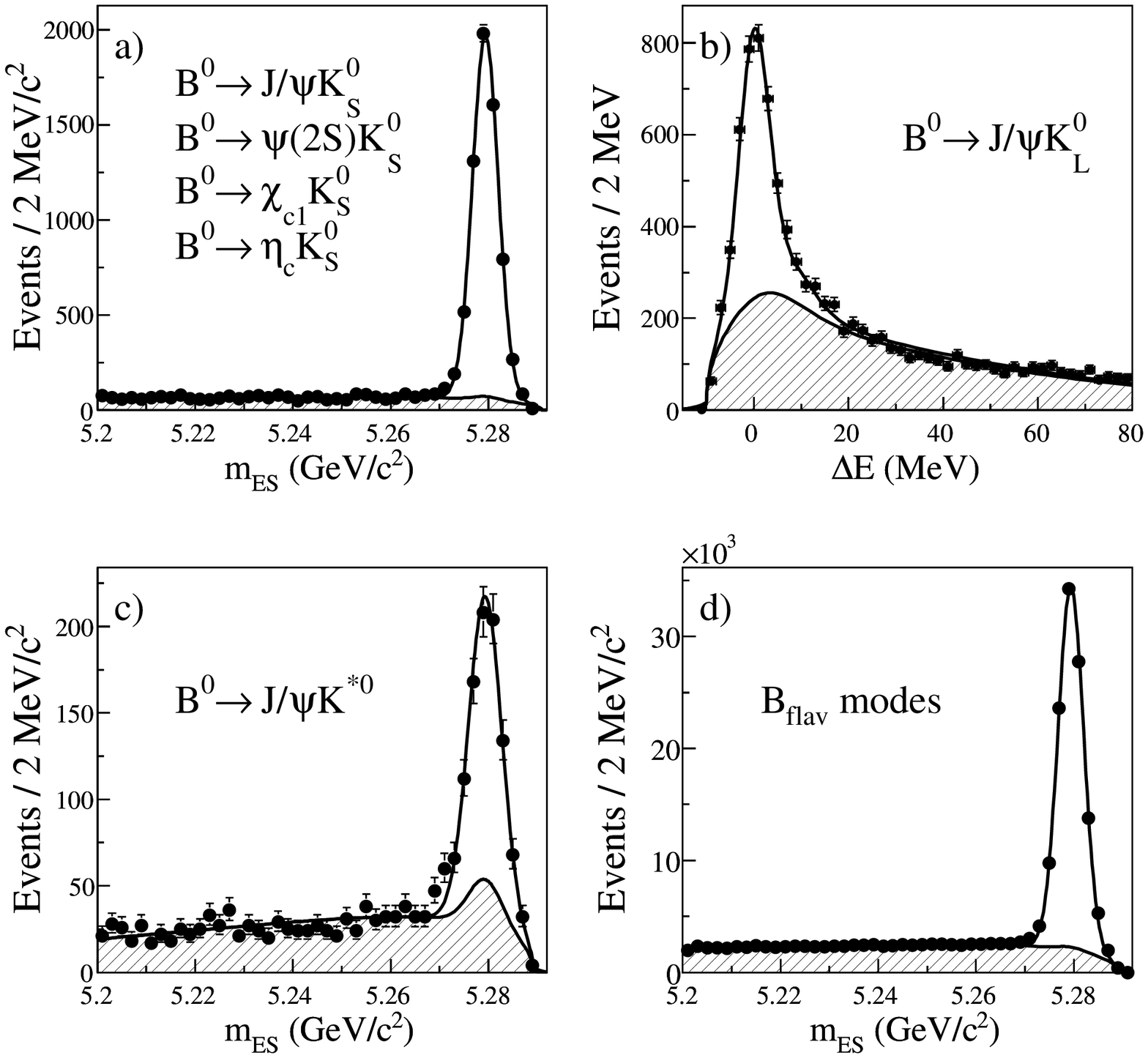}
\caption{
Distributions for \BCP and \Bflav candidates satisfying the tagging and
vertexing requirements:
a) \mes for the final states $\jpsi\KS $, $\psitwos\KS$, $\chicone\KS$,
   and $\etac\KS$,
b) \deltaE for the final state $\jpsi\KL$,
c) \mes for $\jpsi\Kstarz(\Kstarz\to \KS\piz)$, and 
d) \mes for the \Bflav sample. In each plot, the shaded region is the
   estimated background contribution.
}
\label{fig:bcpsample}
\end{center}
\vskip-0.8truecm
\end{figure}

We determine the composition of
our final sample using the beam-energy substituted mass
$\mes=\sqrt{(E^*_\mathrm{beam})^2-(p^*_B)^2}$, where $E^*_\mathrm{beam}$
and $p_B^*$ are the beam energy and \B momentum in the $\epem$
center-of-mass (CM) frame. For the $\jpsi\KL$ mode we instead use the
difference \deltaE between the candidate CM energy and $E^*_\mathrm{beam}$. 
The composition of our final sample is shown in Fig.~\ref{fig:bcpsample}.
We use events with $\mes > 5.2 \gevcc$ ($|\deltaE| < 80\mev$ for $\jpsi \KL$) 
to determine the properties of the background contributions.
We define
a signal region $5.27 < \mes < 5.29 \gevcc$ ($|\deltaE|< 10\mev$ for $\jpsi
\KL$), which contains 12677 \CP candidate events that satisfy the tagging
and vertexing requirements (see Table~\ref{tab:result}).
For all modes except $\etac \KS$ and $\jpsi\KL$, we use simulated events to
estimate the fractions of events that peak in the \mes signal region due
to cross-feed from other decay modes (peaking background). 
For the $\etac\KS$ mode, the cross-feed fraction is determined
from a fit to the $m_{KK\pi}$ and \mes distributions in data.
For the $\jpsi\KL$ decay mode, the sample composition, effective $\eta_f$,
and \deltaE distribution of the individual background sources are 
determined either from simulation (for $B\to\jpsi X$) or from the
$m_{\ellell}$ sidebands in data (for non-\jpsi background).

We determine \stwob and \abslambda from a simultaneous maximum likelihood
fit to the \deltat distribution of the tagged \BCP and \Bflav samples. The
\deltat distributions of the \BCP sample are modeled by
Eq.~\ref{eq:timedist}. Those of the \Bflav sample evolve according to
Eq.~\ref{eq:timedist} with $\lambda=0$.
The observed amplitudes for the \CP asymmetry in the \BCP
sample and for flavor oscillation in the \Bflav\ sample are reduced by the
same factor, $1-2\mistag$, due to flavor mistags.
The \deltat distributions for the signal are convolved with a
resolution function common to both the \Bflav and \BCP samples,
modeled by the sum of three Gaussian functions~\cite{ref:bigprd}.
The combinatorial background is incorporated with an empirical description
of its \deltat spectra, containing prompt and non-prompt lifetime
components convolved with a resolution function~\cite{ref:bigprd} distinct
from that of the signal. The peaking background is assigned
the same \deltat distribution as the signal but with no \CP violation,
with the same \deltat resolution function.

\begin{table}[!htb] 
\vskip-0.4truecm
\begin{center}
\caption{
Number of events $N_{\rm tag}$ and signal purity $P$ in the signal region
after tagging and vertexing requirements, and results of fitting for \CP
asymmetries in the \BCP sample and various subsamples.
In addition, fit results for the \Bflav and $B^+$ control samples demonstrate that
no artificial \CP asymmetry is found where we expect no \CP violation
($\stwob=0$, $\abslambda=1$).
Errors are statistical only.
}
\label{tab:result}
\begin{tabular}{lrccc}\tbline\tbline
Sample  & $N_{tag}$ & \!$P(\%)$\! & $\stwob$ & \abslambda \\ \tbline
Full \CP sample & 12677 & 75 & ~$0.714 \pm 0.032$~ & ~$0.952 \pm 0.022$ \\ \tbline
\JpsiKsCh       &  4459 & 96 & ~$0.702 \pm 0.042$~ & ~$0.976 \pm 0.030$ \\
\JpsiKszz       &  1086 & 88 & ~$0.617 \pm 0.103$~ & ~$0.812 \pm 0.058$ \\
$\psitwos\KS$   &   687 & 83 & ~$0.947 \pm 0.112$~ & ~$0.867 \pm 0.079$ \\
$\chicone\KS$   &   313 & 89 & ~$0.759 \pm 0.170$~ & ~$0.804 \pm 0.102$ \\
$\etac\KS $     &   328 & 69 & ~$0.778 \pm 0.195$~ & ~$0.948 \pm 0.141$ \\
$\jpsi\KL$      &  4748 & 55 & ~$0.734 \pm 0.074$~ & ~$1.061 \pm 0.063$ \\
$\jpsi\Kstarz$  &  1056 & 66 & ~$0.477 \pm 0.271$~ & ~$0.954 \pm 0.083$ \\ \tbline
$\jpsi\Kz$      & 10275 & 76 & ~$0.697 \pm 0.035$~ & ~$0.966 \pm 0.025$ \\ \tbline
$\jpsi\KS$      &  5547 & 94 & ~$0.686 \pm 0.039$~ & ~$0.950 \pm 0.027$ \\ \tbline
$\eta_f=-1$     &  6873 & 92 & ~$0.711 \pm 0.036$~ & ~$0.935 \pm 0.024$ \\ \tbline\hline
1999-2002 data  &  3084 & 79 & ~$0.735 \pm 0.063$~ & ~$0.987 \pm 0.045$ \\
2003-2004 data  &  4850 & 77 & ~$0.728 \pm 0.052$~ & ~$0.940 \pm 0.035$ \\
2005-2006 data  &  4725 & 74 & ~$0.681 \pm 0.054$~ & ~$0.940 \pm 0.037$ \\ \hline\hline
\leptontag      &  1349 & 80 & ~$0.728 \pm 0.066$~ & ~$0.901 \pm 0.043$ \\
\kaonitag       &  1843 & 76 & ~$0.689 \pm 0.063$~ & ~$0.986 \pm 0.046$ \\
\kaoniitag      &  2948 & 72 & ~$0.751 \pm 0.071$~ & ~$0.880 \pm 0.044$ \\
\kpitag         &  2321 & 73 & ~$0.654 \pm 0.112$~ & ~$0.999 \pm 0.075$ \\
\piontag        &  2551 & 76 & ~$0.671 \pm 0.167$~ & ~$0.927 \pm 0.104$ \\
\othertag       &  1665 & 73 & ~$0.705 \pm 0.504$~ & ~$1.506 \pm 0.483$ \\ \tbline\tbline
\Bflav sample   &123893 & 85 & ~$0.018 \pm 0.010$~ & ~$0.995 \pm 0.007$ \\
$B^+$ sample    & 29598 & 94 & ~$0.012 \pm 0.017$~ & ~$1.010 \pm 0.012$ \\\tbline\hline
\end{tabular}
\end{center}
\vskip-0.7truecm
\end{table}
In addition to \stwob and \abslambda, there are 68 free parameters in the
\CP fit. For the signal, these are the parameters of the \deltat
resolution (7), the average mistag fractions \mistag and the differences
$\Delta\mistag$ between \Bz and \Bzb mistag fractions for each tagging
category (12), and the difference between \Bz and \Bzb reconstruction and
tagging efficiencies (7).
The background is described by mistag fractions (24), parameters
of the \deltat resolution (3) and \Bflav time dependence (3), and 
parameters for the \CP background (8), including the
apparent \CP asymmetry of non-peaking events in each tagging category.
Finally, we allow for the possibility of direct \CP violation in
the $\chicone\KS$ background to $\jpsi\Kstarz$ (1), and in the main
backgrounds to the $\jpsi\KL$ mode, coming from $\jpsi\KS$, $\jpsi\Kstarz$, and
the remaining $\jpsi$ background (3 parameters). The effective \abslambda
of the non-\jpsi background is fixed from a fit to the \jpsi-candidate
sidebands in $\jpsi\KL$. We fix $\tau_{\Bz}=1.530\ps$ and
$\deltamd=0.507\ps^{-1}$~\cite{ref:pdg2006}.
The determination of the mistag fractions and \deltat resolution
function parameters for the signal is dominated by the
\Bflav sample, about 10 times more abundant than the \CP sample.

The fit to the \BCP and \Bflav samples yields $\stwob = \fitstwobstat$ and
$\abslambda = \fitlambdastat$, where the errors are statistical only.
The correlation between these two parameters is $-1.5\,\%$. We also
perform a separate fit in which we allow different \stwob and \abslambda
values for each charmonium decay mode, a fit to the
$\jpsi\KS\,(\pipi+\ppz)$ mode, and a fit to the $\jpsi\Kz\,(\KS+\KL)$
sample. We split the data sample by run period and by tagging 
category. We perform the \CP measurements on control samples with 
no expected \CP asymmetry. The results of these fits are summarized in
Table~\ref{tab:result}. The difference in the $\etac \KS$ \stwob value with
respect to our previous publication~\cite{ref:babar2004} is partly due to
the slightly different reconstruction algorithms and partly to the different
selection; the two measurements are consistent when the systematic error is
taken into account.

\begin{figure}
\begin{center}
\includegraphics*[bb=0 20 567 705,height=10.0cm,clip=]%
{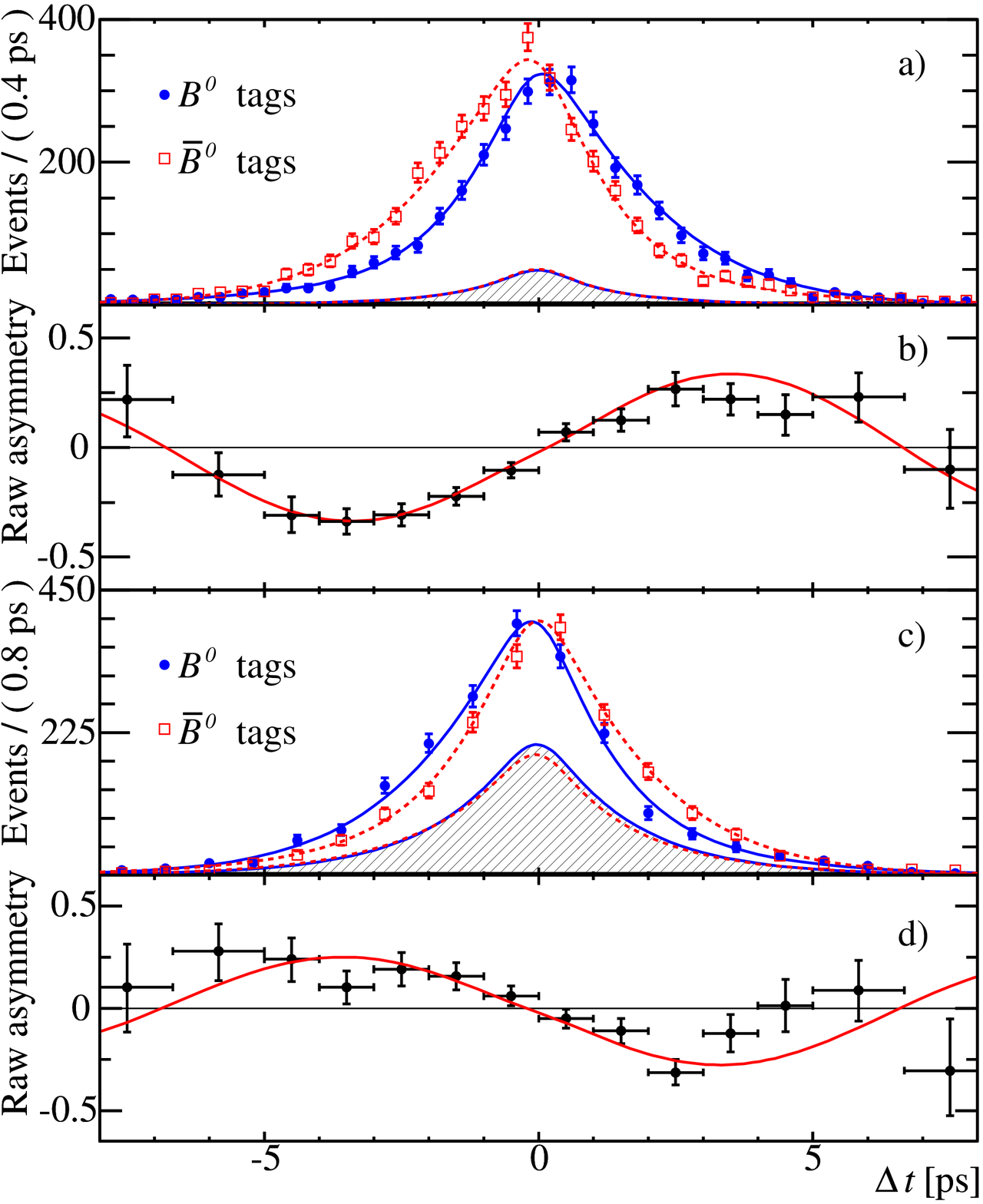}
\caption{
a) Number of $\eta_f=-1$ candidates ($\jpsi\KS$, $\psitwos\KS$,
$\chicone \KS$, and $\etac \KS$) in the signal region with a \Bz tag
($N_{\Bz }$) and with a \Bzb tag ($N_{\Bzb}$), and
b) the raw asymmetry, $(N_{\Bz}-N_{\Bzb})/(N_{\Bz}+N_{\Bzb})$, as functions
of \deltat.
Figures c) and d) are the corresponding distributions for the $\eta_f=+1$
mode $\jpsi\KL$.
To enhance the signal component, all distributions exclude \othertag-tagged events.
The solid (dashed) curves represent the fit projections in \deltat for \Bz
(\Bzb) tags. The shaded regions represent the estimated background
contributions.}
\label{fig:cpdeltat}
\end{center}
\vskip-1.0truecm
\end{figure}

Figure~\ref{fig:cpdeltat} shows the \deltat distributions and 
asymmetries in yields between events with \Bz tags and \Bzb tags for the
$\eta_f=-1$ and $\eta_f = +1$ samples as a function of \deltat,
overlaid with the projection of the likelihood fit result.
We also performed the \CP fit fixing $\abslambda=1$, which yields
$\stwob=0.713\pm0.032\,\stat$.

The dominant systematic errors on \stwob are due to limited
knowledge of various background properties, incuding uncertainties in
$\jpsi\KL$-specific backrounds and in the amounts of peaking backgrounds
and their \CP asymmetries (0.010), to possible differences between the
\Bflav and \BCP tagging performances (0.009), to the description of the
\deltat resolution functions (0.008), to the knowledge of the
event-by-event beamspot position (0.005).
The only sizeable sysyematic uncertainties on \abslambda are due to the 
possible interference between the suppressed $\bar b\to \bar u c \bar
d$ amplitude with the favored $b\to c \bar u d$ amplitude for some tag-side
\B decays~\cite{ref:dcsd} (0.015), and to the \CP content of the peaking
backgrounds (0.006).
The total systematic error on \stwob (\abslambda) is \syststwob\
(\systlambda).
The main systematic uncertainties on both \stwob and \abslambda for the full 
sample, for the seven individual modes, and for the fits to the $\jpsi\Kz$
and $\jpsi\KS$ samples are summarized in the Table
at~\cite{ref:EPAPS_systematicssplitbymode}.

The large \BCP sample allows a number of consistency checks, including
separation of the data by decay mode and tagging category.
The results of those checks, all consistent within the errors, are listed
in Table~\ref{tab:result}.
We observe no statistically significant asymmetry
from fits to the control samples of non-\CP decay modes.

In summary, we report improved measurements of \stwob and \abslambda
that supersede our previous results~\cite{ref:babar2004}.
We measure $\stwob=\fitstwob\pm\statstwob\,\stat\pm \syststwob\,\syst$ and
$\abslambda=\fitlambda\pm\statlambda\,\stat\pm\systlambda\,\syst$,
providing an improved model-independent constraint on the position of
the apex of the unitarity triangle~\cite{ref:CKMfits}.
Our measurements agree within errors with the published
results~\cite{ref:HFAG2006,ref:belle} and with the theoretical estimates of
the magnitudes of CKM matrix elements in the context of the
SM~\cite{ref:CKMestimates}.
The measured value of \abslambda is consistent with no direct \CP violation
with a significance of 1.72 standard deviations.
We report the first individual measurements of \stwob and \abslambda for
each of the decay modes within our \CP sample, and of the
$\jpsi\Kz$($\KS+\KL$) sample.

We are grateful for the excellent luminosity and machine conditions
provided by our \pep2\ colleagues, 
and for the substantial dedicated effort from
the computing organizations that support \babar.
The collaborating institutions wish to thank 
SLAC for its support and kind hospitality. 
This work is supported by
DOE
and NSF (USA),
NSERC (Canada),
IHEP (China),
CEA and
CNRS-IN2P3
(France),
BMBF and DFG
(Germany),
INFN (Italy),
FOM (The Netherlands),
NFR (Norway),
MIST (Russia),
MEC (Spain), and
PPARC (United Kingdom). 
Individuals have received support from the
Marie Curie EIF (European Union) and
the A.~P.~Sloan Foundation.



\begin{thebibliography}{99}
\bibitem{ref:CKM}
N.~Cabibbo, Phys.~Rev.~Lett.~{\bf 10}, 531 (1963);
M.~Koba\-yashi and T.~Maskawa, Prog.\ Th.\ Phys.\ {\bf 49}, 652 (1973).

\bibitem{BCP}
A.B. Carter and A.I. Sanda, \jprd{23}, 1567 (1981);
I.I. Bigi and A.I. Sanda, \np {\bf 193}, 85 (1981).

\bibitem{ref:lambda}
See, for example, D.~Kirkby and Y.~Nir in Ref.~\cite{ref:pdg2006}.

\bibitem{ref:chargeconj}
Charge-conjugate reactions are implicitly included.

\bibitem{ref:babar2004}
\babar\ Collaboration, B.\ Aubert \etal,
\jprl{94}, 161803 (2005).

\bibitem{ref:grossman}
Y. Grossman \etal, \jpl{B538}, 327 (2002).

\bibitem{ref:pdg2006}
W.-M.~Yao \etal, (Particle Data Group), \jpg{33}, 1 (2006).

\bibitem{ref:babar}
\babar\ Collaboration, B.\ Aubert \etal,
\nima{479}, 1 (2002).

\bibitem{ref:rperp}
\babar\ Collaboration, B.\ Aubert \etal, arXiv:0704.0522v1 [hep-ex]
[Phys. Rev. D (to be published)].

\bibitem{ref:bigprd}
\babar\ Collaboration, B.\ Aubert \etal,
\jprd{66}, 032003 (2002).

\bibitem{ref:etacks} 
\babar\ Collaboration, B.\ Aubert \etal, 
\jprd{70}, 011101 (2004).

\bibitem{ref:dcsd}
O.~Long, M.~Baak, R.~N.~Cahn, and D.~Kirkby,
\jprd{68}, 034010 (2003).

\bibitem{ref:EPAPS_systematicssplitbymode}
See EPAPS Document No. E-PRLTAO-99-075743 for a table with the main systematic
uncertainties on \stwob and \abslambda for the full sample, for the seven
individual modes, and for the fits to the $\jpsi\Kz$ and $\jpsi\KS$
samples. For more imformation on EPAPS, see
\texttt{http://www.aip.org/pubservs/epaps.html}.

\bibitem{ref:CKMfits}
UTfit Collaboration (M. Bona \etal), \jprl{97}, 151803 (2006);
CKMfitter Group (J. Charles \etal), \epjc{41}, 1-131 (2005).

\bibitem{ref:HFAG2006}
E.~Barberio {\it et al.} (Heavy Flavor Averaging Group), arXiv:hep-ex/0603003.

\bibitem{ref:belle}
Belle Collaboration, K.-F.\ Chen \etal, \jprl{98}, 031802 (2007).

\bibitem{ref:CKMestimates}
M.~Ciuchini \etal, Z.\ Phys.\  C {\bf 68}, 239 (1995).

\end{thebibliography}
\end{document}